\documentclass[conference]{IEEEtran}

\usepackage{hyperref}

\usepackage[latin1]{inputenc}
\usepackage{chngcntr}
\usepackage[cmex10]{amsmath}
\usepackage[T1]{fontenc}
\usepackage{lettrine}
\usepackage{amsfonts}
\usepackage{amssymb}
\usepackage{graphicx}
\usepackage{csquotes}
\usepackage{mathtools}
\usepackage{bbm}
\usepackage[normalem]{ulem}
\usepackage{tabularx}
\usepackage{float}
\usepackage{enumerate}
\usepackage[numbers,sort&compress]{natbib}
\usepackage[dvipsnames]{xcolor}
\usepackage{tikz}
\usepackage{physics}
\usepackage{todonotes}

\usepackage{algorithm}
\usepackage[noend]{algpseudocode}

\usepackage[final]{microtype}

\usetikzlibrary{arrows.meta}

\usepackage{algorithm}
\usepackage{algpseudocode}
\algnewcommand\algorithmicforeach{\textbf{for each}}
\algdef{S}[FOR]{ForEach}[1]{\algorithmicforeach\ #1\ \algorithmicdo}

\tikzset{>=latex}

\algnewcommand{\LeftComment}[1]{\Statex \(\triangleright\) #1}

\makeatletter
\renewcommand{\ALG@name}{Protocol}
\renewcommand{\complement}{{\mathsf{c}}}
\makeatother
\algblock{Input}{EndInput}
\algnotext{EndInput}
\algblock{Output}{EndOutput}
\algnotext{EndOutput}
\algnotext{EndIf}
\algrenewcommand{\algorithmicthen}{:}

\usetikzlibrary{arrows,
                positioning,
                decorations.pathmorphing,
                decorations.markings,
                decorations.pathreplacing,
                shapes,
                fadings,
                calc
            }

\counterwithin{equation}{section}
\newcounter{thm}
\newtheorem{theorem}[thm]{Theorem}

\newtheorem{remark}[thm]{Remark}
\newtheorem{definition}[thm]{Definition}

\newcommand{\ciel}[1]{\lceil #1 \rceil}

\newcommand*{\rom}[1]{\uppercase\expandafter{\romannumeral #1\relax}}

\newcommand{\eps}{{\varepsilon}}        

\newcommand{\eins}{{\mathbbm{1}}}

\allowdisplaybreaks

\makeatletter
\newcommand*{\algrule}[1][\algorithmicindent]{%
  \makebox[#1][l]{%
    \hspace*{.2em}
    \vrule height .75\baselineskip depth .25\baselineskip
  }
}

\newcount\ALG@printindent@tempcnta
\def\ALG@printindent{%
    \ifnum \theALG@nested>0
    \ifx\ALG@text\ALG@x@notext
    \else
    \unskip
    \ALG@printindent@tempcnta=1
    \loop
    \algrule[\csname ALG@ind@\the\ALG@printindent@tempcnta\endcsname]%
    \advance \ALG@printindent@tempcnta 1
    \ifnum \ALG@printindent@tempcnta<\numexpr\theALG@nested+1\relax
    \repeat
    \fi
    \fi
}

\newcounter{protocol}
\newenvironment{protocol}[1][htb]{%
  \let\c@algorithm\c@protocol
  \renewcommand{\ALG@name}{Protocol}
  \begin{algorithm}[#1]%
  }{\end{algorithm}
}

\counterwithout{equation}{section} 

\title{Undoing Causal Effects of a Causal Broadcast Channel with Cooperating Receivers using Entanglement Resources\thanks{This work was financed by the DFG via grant NO 1129/2-1.}}

\author{
\IEEEauthorblockN{Stephen DiAdamo, Janis N\"otzel }
\IEEEauthorblockA{\textit{Emmy-Noether Gruppe Theoretisches Quantensystemdesign}\\\textit{Lehrstuhl f\"ur Theoretische Informationstechnik} \\
\textit{Technische Universit\"at M\"unchen} \\
\{stephen.diadamo, janis.noetzel\}@tum.de}}

\begin{document}
\maketitle

\thispagestyle{plain}
\pagestyle{plain}

\renewcommand{\complement}{\mathsf{c}}
\let\oldsout\sout
\renewcommand{\sout}[1]{\textcolor{red}{\oldsout{#1}}}
\newcommand{\red}[1]{\textcolor{red}{\textbf{#1}}}
\newcommand{\green}[1]{\textcolor{green}{\textbf{#1}}}
\newcommand{\jn}[1]{\textcolor{violet}{\textbf{#1}}}

\begin{abstract}
	We analyse a communication scenario over a particular causal broadcast channel whose state depends on a modulo sum. The receivers of the broadcast receive channel state information and collaborate to determine the channel state as to decode their private messages. Further, the receivers of the broadcast can collude up to the minimum non-collusion condition to determine state information of the other non-colluding receivers. We analyse three resource scenarios for the receivers: receivers can share entanglement without classically communicating, can just use classical communication, or have both entanglement and classical communication. Using results from secure multi-party communication, we find that when the receivers can share entanglement and communicate classically, they can receive messages from the sender at a non-zero rate with verifiable secure collaboration. In the entanglement only case a positive capacity is not possible. In the classical communication case, a non-zero rate of communication is achievable but the communication complexity overhead grows quadratically in the number of receivers versus linear in the number of receivers with entanglement.
\end{abstract}


\section{Introduction}\label{sec:introduction}

\lettrine[lines=2, nindent=0em, lraise=0.1]{T}{}he use of entanglement resources in classical communication scenarios has been shown to reduce communication resources \cite{gavinsky2006role}, increase capacity of channels \cite{bennett1999entanglement}, and enable communication \cite{notzel2020entanglement, noetzel2020internetofthings}. Entanglement can moreover be used to improve upon protocols for secure multi-party computation and communication  \cite{crepeau2002secure, ekert1991quantum}, allowing parties to compute a common function securely. In this work, we introduce a communication scenario with a single sender broadcasting classically encoded messages to $N$ receivers. To decode the messages, the receivers must cooperate, or conference \cite{maric2007capacity}, with one another or otherwise, because of the channel's causal properties, the capacity of the channel vanishes. This scenario mimics one where a sender  broadcasts a message to a collection of receivers such that either all receivers receive the message or none of them receive it, based on their mutual cooperation.

The channel model in this work has a causal state variable that determines how the channel transmits the sender's messages. Specifically \enquote{causal} here means that with each transmission made over the channel, the transmission behaviour of the channel changes depending on a random state variable. When a message arrives at the receiver, it is accompanied with a piece of the state information. The state information is split between the receivers such that alone, the best the receiver can do is randomly guess the channel state. The receivers' task is to cooperate to fairly determine the channel state -- essential for decoding the sender's message. In the general case, the receivers can collude with each other in groups to attempt to determine the channel state unfairly, that is, where the receivers in the colluding group determine the channel state and the others do not. To overcome this, we apply secure multi-party protocols that work under a maximum colluding number of receivers. We investigate protocols under three conferencing resource scenarios.

The scenarios we consider are: when the receivers are only able to share entanglement with each other, but are unable to classically communicate; second, they can classically communicate but not share entanglement; and lastly when they can both share entanglement and can classically communicate. In these scenarios, we determine a lower-bound for the communication cost for performing protocols that allow the receivers to fairly determine the channel state. We find that with only classical resources, the receivers can perform a fair and conditionally secure protocol to determine the channel state but at the expense of additional classical resources compared to the case where they can in addition share entanglement. We see an overall quadratic reduction of the communication resources needed when entanglement resources are available with an addition of provable security.

\subsection{Related Work}

In this work we consider a causal broadcast channel communication scenario with cooperating receivers. In \cite{noetzel2020internetofthings}, a dual scenario is considered where instead of a cause broadcast channel, the channel is a causal multiple access channel with cooperating senders. In that case, the receivers need not communicate, as shared entanglement is enough to achieve a positive channel capacity. In
\cite{pereg2019arbitrarily}, a causal broadcast channel with channel state information at the sender is analysed.  In this work, we make use of the multi-party modulo summation protocols developed in \cite{chor1993communication} and \cite{hayashi2019verifiable}, extending and optimising them for this this particular communication scenario when $n$ modulo sums are needed.

\subsection{Summary of Contributions}

We propose a communication model with a sender and many receivers that can be used to ensure that either all receivers receive their message with certainty if they cooperate or otherwise none of them receive their message. We use protocols developed in \cite{chor1993communication, hayashi2019verifiable} to develop a scheme that is fair and secure such that all receivers can decode messages from the sender fairly when the protocols run honestly, even when some receivers collude. We consider the trade-offs for allowing the receivers to share classical and quantum resources, and find when the receivers can share entanglement and can broadcast messages, they can verifiably securely perform the multi-party computation needed to determine the channel state which encrypts each transmission from the sender. We determine the communication complexity for each case where communication is possible. 

\subsection{Notation}\label{sec:notation}
 Given a finite alphabet $\mathbf X$, the set of probability distributions on it is $\mathcal P(\mathbf X)$. The corresponding state space for quantum systems on a finite dimensional Hilbert space $\mathcal H$ is denoted $\mathcal S(\mathcal H)$. The $n$-fold composition $\mathbf X\times\ldots\times\mathbf X$ is written $\mathbf X^n$. A classical channel $W$ with input alphabet $\mathbf X$ and output alphabet $\mathbf Y$ is defined by a matrix $(w(y|x))_{x\in\mathbf X,y\in\mathbf Y}$ where $w(\cdot|x)\in\mathcal P(\mathbf Y)$ for all $x\in\mathbf X$. The set of all such channels is denoted $C(\mathbf X,\mathbf Y)$. Specific channels using in this work are: The identity map on $x \in \{0,1\}$, defined by $\mathbbm1(x) = x$, the bit-flip on $\mathbbm F$ as $\mathbbm F(x)=x\oplus 1$, and the binary symmetric channel (BSC) with parameter $\nu\in[0,1]$ defined as $BSC(\nu)\coloneqq\nu\mathbbm1+(1-\nu)\mathbbm F.$ The entropy of $p\in\mathcal P(\mathbf X)$ is $H(p)\coloneqq-\sum_{x\in\mathbf X}p(x)\log(x)$, ($\log$ being calculated with base $2$ and using the convention $0\log(0)=0$). The mutual information of $p\in\mathcal P(\mathbf X)$ and $W\in C(\mathbf X,\mathbf Y)$ is $I(p;W) \coloneqq H(p)+H(Wp)-H((W,p))$. Distribution $\pi \in \mathcal P(X)$ represents the uniform distribution on $X$.

\section{Channel Model}

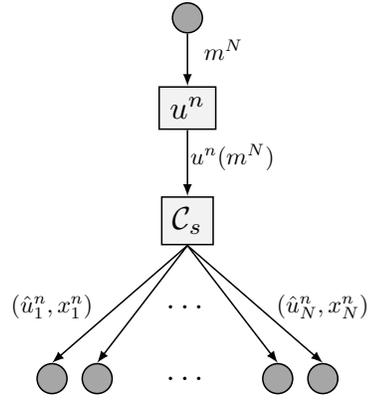
\begin{figure}
	\centering
	\begin{tikzpicture}[scale=1.2, every node/.style={transform shape}]
		\node[circle, draw, fill=gray!70, line width=0.2mm] (s) at (0, 0) {};
		\node[circle, draw, fill=gray!70, line width=0.2mm] (x1) at (-1.5, -4) {};
		\node[circle, draw, fill=gray!70, line width=0.2mm] (x2) at (-1, -4) {};
		\node[circle, draw, fill=gray!70, line width=0.2mm] (x3) at (1, -4) {};
		\node[circle, draw, fill=gray!70, line width=0.2mm] (x4) at (1.5, -4) {};
		\node[draw, fill=gray!10, line width=0.2mm] (u) at (0, -1) {$u^n$};
		    
		\node[scale=0.75] () at (.4, -.35) {$m^N$};
		\node[scale=0.75] () at (.5, -1.55) {$u^{n}(m^N)$};
		\node[scale=0.75] () at (-1.5, -3.2) {$(\hat{u}^{n}_1, x^n_1)$};
		\node[scale=0.75] () at (1.5, -3.2) {$(\hat{u}^{n}_N, x^n_N)$};
		    
	    \node[] () at (0, -3.2) {$\dots$};
		\node[] () at (0, -4) {$\dots$};
		    
		\node[draw, fill=gray!10, line width=0.2mm] (c) at (0, -2.25) {$\mathcal{C}_s$};
		    
		\draw[->, line width=.2mm] (s.south) -- (u.north);
		\draw[->, line width=.2mm] (u.south) -- (c.north);
		\foreach \y in {1, ..., 4}{
			\draw[->, line width=.2mm] (c.south) -- (x\y.north);
		}
		    
	\end{tikzpicture}
	\caption{Depiction of the causal, state dependent, broadcast channel. The sender at the top sends $m^N$ to the encoder $u^n$ which encodes the message as $u^{Nn}$. After encoding, the message is put through the state dependent channel $\mathcal{C}_s$, where for each bit of $u^{Nn}$ sent through the channel, a state is selected uniformly at random game using the modulo sum of $x^N$. One piece of $x^{N}$ is sent to the respective receiver until all $x_i^n$ are received. The receivers can then use their part of $x^{Nn}$ as input to their coordinator $q$ to receive $\gamma_i^n$ which can be used to aid in decoding $\hat{u}_i^n$.}
	\label{fig:broadcast_chan}
\end{figure}

The broadcast channel consists of one sender and  $N$ receivers, of which up to $N-2$ can be colluding, as to satisfy maximum collusion condition. A single sender intends to transmit a collection of private messages $m^N \coloneqq (m_1, ..., m_N)$, one to each of the respective receivers. For message transmission, the sender uses encoder $u^n: (m_1,...,m_N)\mapsto (u_1^n,...,u_N^n)$ which maps the messages $(m_1,..., m_N)$ to block-length $n$ messages $u_i^n \in \{0,1\}^{Nn}$ for $n$ channel uses. For convenience we define the map $u^n(m_i) \mapsto u^n(m^N)_i$, a mapping to the $i$-th index of the encoded message string. Between the sender and each receiver $i$ is a state dependent binary inference channel $\mathcal{C}_{s} \in C(\{0,1\}, \{0, 1\})$ with an environmentally controlled state $s\in\{0,1\}$. The channel behaves as follows: When $s=0$, $\mathcal{C}_{0} \coloneqq \mathbbm{1}$ is the identity channel, and when $s=1$, $\mathcal{C}_{1} \coloneqq \mathbbm{F}$ is the bit flip channel. The total channel $\mathcal{C}_s^N$ is a product of the channels, that is, 
\begin{align}
	\mathcal{C}_s^N \coloneqq \prod_{i=1}^N \mathcal{C}_{s}, \label{eq:product_chan} 
\end{align}
where each channel in the product shares the same state $s$.

The main requirement for channel state selection is that the state of the channel is uniformly random between two outcomes such that for three or more players, no receiver can infer the channel state based on their individual state information. We use one such example as follows. Prior to transmission of $u_t=\{0,1\}^N$ at time $t$, an environmental state $x_t^N \in \{0, 1\}^N$ is realized according to the uniform distribution on $\{0, 1\}^N$. For each $i$, One part $x_{i,t}$ of $x^N_t$ is sent along with other information to receiver $i$ with no encoding or effects from the channel when a transmission is made. The state $s$ at transmission $t$ is selected using  $s = \sum_{i} x_{i,t} \mod 2$. For convenience we define $\Phi(x_t^N) = \sum_{i} x_{i,t} \mod 2$. We use this example modulo sum throughout this work. $\Phi(x^{Nn})$ denotes $(\Phi(x^N_1), ..., \Phi(x^N_n))$.

When the sender sends a message to the receivers, the message is encoded into $n$ bits and transmitted over $n$ uses of $\mathcal{C}_s^N$. At each transmission $t$ the receivers receive their part of $x^N$ and use it to assist in decoding their private message from the sender. Each receiver $i\in[N]$ can use $\{x_{i,t}\}_{t=1}^n$ to aid in selecting a decoder $d^n_i \in C(\{0, 1\}^n \times \Gamma^n, \mathbf{M}_i)$, $\mathbf{M}_i$ the codebook for receiver $i$, which takes as input $n$ outputs of the channel $\mathcal{C}_{s}$ and a coordination parameter $\gamma_i^n \in \Gamma^n$ which serves to coordinate the $N$ receivers while decoding. The parameter $\gamma_{i}^n\in\Gamma^n$ is distributed according to the channel $q \in C(\{0,1\}^{Nn}, \Gamma^{Nn})$ which we call the coordinator. Each receiver inputs their part $x_{i}^n$ of $x^{Nn}$ to $q$ and receives a response $\gamma_{i}^n$ which is used for decoding. 

\begin{definition}[Non-signalling channel]
    A channel $q \in C(\mathbf{X}, \Gamma)$ is called non-signalling if, for all $\gamma_1, x_1, x_2, x_2'$,
    \begin{align}
        &\sum_{\gamma_2} q(\gamma_1, \gamma_2|x_1, x_2) = \sum_{\gamma_2} q(\gamma_1, \gamma_2|x_1, x_2')
    \end{align}
    and for all $\gamma_2, x_1, x_1', x_2$,
    \begin{align}
        &\sum_{\gamma_1} q(\gamma_1, \gamma_2|x_1, x_2) = \sum_{\gamma_1} q(\gamma_1, \gamma_2|x_1', x_2).
    \end{align}
\end{definition}

In this work, we use various approaches at the receivers for determining the decoding parameter $\gamma^n_{i}$. Here we consider the following three scenarios:

\begin{enumerate}
    \item Classical communication is not available between receivers, but shared entanglement is.
    \item Classical communication is available between receivers but not entanglement.
    \item Classical communication and shared entanglement are both available.
\end{enumerate}
In the first case, correlations are non-signalling, where in the other two, they are signalling.
We define the channel formally as follows:

\begin{definition}[Broadcast Code]\label{def:broadcast_code}
	A broadcast code $C$ for block-length $n$  and $N$ receivers consists of message sets $\mathbf{M}_1, ..., \mathbf{M}_N$. Moreover $C$ contains an encoder $u^n(m_1, ..., m_N)$ where $u^n : (m_1, ..., m_N) \mapsto (u_1^n, ..., u_N^{n}) \subset \{0,1\}^{Nn}$. Next, for each $i$, $C$ contains a causal decoder $d^n_{i}(\hat{m}_i |u^n_{i}, \gamma_{i}^n)$ which decodes the channel output when $n$ bits arrive. $d_i^n$ assigns an estimate $\hat{m}_i \in \mathbf{M}_i$ using decoding sets $D^{\gamma_i}_i\coloneqq\{D^{\gamma_i}_{i,j}\}_{j=1}^{|\mathbf{M}_i|} \subset \{0,1\}^n$ where for a fixed $i$ and $j\neq k$ $D^{\gamma_i}_{i,j} \cap D^{\gamma_i}_{i, k} = \emptyset$. 
\end{definition}

\begin{remark}
$D^{\gamma_i}_i$ is dependent on $\gamma_i$ where the dependence is to counteract the estimated the channel state.
\end{remark}

\begin{definition}[Transmission Error]
	Let $C_n$ be a broadcast code with block-length $n$ and $N$ receivers transmitting over $\mathcal{C}_s^N$ as defined in \eqref{eq:product_chan}. Let $M \coloneqq \prod_{i=1}^N |\mathbf{M}_i|$. The average probability for transmission success is, 
    \begin{align}
        \begin{aligned} 
        P_{succ}(C_n) = \frac{1}{M 2^{Nn}} \sum_{m^N}\sum_{\substack{x^{Nn}, \\ \gamma^{Nn}}}  & q(\gamma^{Nn}|x^{Nn}) \cdot  \\[-3mm] 
        & \hspace{-18mm}   \prod_{i=1}^N d^n_{i}(D^{\gamma_i^n}_{i,m_i}| \mathcal{C}^n_{\Phi(x^{Nn})}(u^n(m_i)), \gamma_{i}^n).
        \end{aligned}
    \end{align}
	where $\mathcal{C}^n_{\Phi(x^{Nn})}(u^n(m))$ represents the channel output after $n$ uses of $\mathcal{C}_{\Phi(x_t^N)}$ using $\{x^N_t\}_{t=1}^n$ for transmission $t$ to send the $t$-th bit of encoding $u^n(m)$. The probability of error is defined as $P_{err}(C_n) = 1-P_{succ}(C_n)$.
\end{definition}

\begin{definition}[Achievable Rate Tuple]
	A rate tuple $(R_1, ..., R_N)$ is called achievable if there exists a sequence of causal codes $(C_n)_{n\in\mathbb{N}}$ such that for any $i \in \{1,...,N\}$, 
	\begin{align}
		\liminf_{n \rightarrow \infty} \frac{1}{n} \log |\mathbf{M}_{i,n}| \geq R_i 
	\end{align}
	while at the same time
	\begin{align}
		\lim_{n \rightarrow \infty} P_{err}(C_n) = 0. 
	\end{align}
	 The capacity region of a channel is the closure of all achievable rate tuples.
\end{definition}

In order to determine the channel state for each transmission in this communication scenario, the receivers will need a scheme such that they can determine the modulo sum of the combined state information of each other receiver. Since some of the receivers are possibly colluding, each receiver should not simply give away their own state information and thus use a random variable $R_i$ to encode $X_i$ with. Formally, we enforce that for any proposed protocol for multi-party modulo sum, it must be that for colluding receivers $Z \subset [N], |Z| \leq N - 2$ and each for all non-colluding receivers $i \in [N] \setminus  Z$,
\begin{align}
    I(R_i;Y_Z) = 0, \label{cond:no_info}
\end{align}
where $Y_Z$ is all of the information obtainable by the coalition excluding information derivative to the multi-party calculation. One might consider enforcing $I(X_i;Y_z) = 0$ for all $i$ as well, but for any multi-party summation protocol, there is always a way to determine the sum of the inputs of the non-colluding parties amongst the coalition. We also enforce a reliability condition which is that when all $N$ receivers perform the protocol honestly, then each of the receivers determine the channel state. Once the channel state is determined for each transmission, they generate a variable $\gamma^{Nn}$ which contains the state information for each transmission $1\leq t \leq n$. With this, each receiver $i$ can choose the respective decoding set $D^{\gamma_i^n}_i$.

\section{Entanglement Without Classical Communication}
\begin{theorem}\label{thm:no_comms}
When classical communication between the receivers is not available but entanglement is, the rate region equals $0^N$.
Moreover, without classical communication, no protocol exists using only entanglement to compute the modulo sum of the channel state information.
\end{theorem}

In this scenario, the receivers cannot classically communicate but can share entanglement resources in any form distributed prior transmission. For each transmission $t$, each receiver $i$ receives their bit $x_{i,t}$ where based on the channel state selection mechanism, each $x_{j,t}$ is independent of $x_{i,t}$ for $i\neq j$. The task for the receivers is to correlate themselves such that they have a better than $p=1/2$ chance at guessing the channel state. To do this, the parties  need to devise a joint measurement on their entangled states that can signal, contradicting the no-signaling theorem.  Because the channel state is selected using an unbiased modulo sum the channel state equally the identity channel as it is the bit flip channel, the overall channel is a binary symmetric channel with $p=1/2$, well known to have $0$ capacity. That the parties are malevolent makes no difference here as there is no way for the receivers to gain any information from the other receivers.

\begin{IEEEproof}
For the channel $C_s^N$, assume there is an achievable rate tuple $(R_1, ..., R_N)$ where for at least one $i$, $i=1$ without loss of generality, $R_i > 0$. Then, for any $\epsilon > 0$ there is an $n$ large enough such that,
\begin{align}
 &\begin{aligned}\label{eqn:single-code-positive-rate}
 1-\eps \leq  \frac{1}{M2^{Nn}}\sum_{m^N}\sum_{\substack{x^{Nn}\\ \gamma^{Nn}}} q&(\gamma^{Nn}|x^{Nn}) \cdot & \\[-8mm] & \prod_{i=1}^N d^n_{i}(D^{\gamma_i^n}_{i,m_i}| \mathcal{C}^n_{\Phi(x^{Nn})}(u^n(m_i)), \gamma_{i}^n)
 \end{aligned} \\[-3.5mm]
 &\begin{aligned}
 \leq \frac{1}{\mathbf{M}_1 2^{Nn}}\sum_{m_1}\sum_{\substack{x^{Nn}\\ \gamma^{Nn}}} q(\gamma^{Nn}&|x^{Nn}) \cdot \\[-8mm] &  d^n_{1}(D^{\gamma_1^n}_{1,m_1}| \mathcal{C}^n_{\Phi(x^{Nn})}(u^n(m_1)), \gamma_{1}^n).
 \end{aligned}
\end{align}
In order to better predict $\Phi(x^{Nn})$, the receivers coordinate with $q(\gamma^{Nn}|x^{Nn})$. Shared entanglement alone is non-signalling \cite{popescu1998causality} and so
\begin{align}
    &\sum_{\gamma^{(N-1)n}}q(\gamma^n_1,\gamma^{(N-1)n}|x_1^n, x^{(N-1)n})   
    = q_1(\gamma^n_1|x_1^n).
\end{align}

The channel state $\Phi(x^{Nn})$ for channel $\mathcal{C}^n_{\Phi(x^{Nn})}(\hat{u}^n|u^n(m_1))$ and decoder $d_1^n$ with $n$ transmissions leads to an effective channel from sender $1$ to receiver $1$. Let $g$ denote the channel from sender $1$ to receiver $1$, including the reception of a copy $x$ of the state variable $x_1$.
\begin{align}
    g(x,y|u)
        &=2^{-N}\sum_{x^N}\delta(x,x_1)\delta(y,\Phi(x^N)\oplus u)\\
        &=\tfrac{1}{2}\sum_{x_1,\hat x}\delta(x,x_1)\tfrac{1}{2}\delta(y,\hat x\oplus x_1 \oplus u)\\
        &=\pi(x)\pi(y).
\end{align}
Obviously, for the Shannon capacity $C$ of $g$ it holds $C(g)=0$. Therefore no coding scheme, and in particular none that uses the particular random decoder 
\begin{align}
    \hat d(m|x_1^n,y^n):=\sum_{\gamma^n_1}q(\gamma^n_1|x^n_1) \eins_{D^{\gamma^n_1}_{m}}(y^n),
\end{align}
can transmit at positive rates over $g$. This contradicts \eqref{eqn:single-code-positive-rate} and thus only $0^N$ is achievable. 
\end{IEEEproof}

\section{Classical Communication Without Entanglement}
In the next two cases, we develop a signalling coordinator $q$ that can be used to achieve positive capacity.

\begin{theorem}\label{thm:cla_comms}
When classical communication is allowed between the receivers, using Strategy \ref{proto:classic2} the rate tuple $1^N$ is achievable and the communication complexity is $\Omega(N^2)$ using a conditionally secure protocol.
\end{theorem}

\begin{remark}
For unconditional security, the secrecy of each communication channel between each pair of connected receivers must be verified, which would add a significant communication resource overhead to the protocol. With only classical resources, one could consider private key distribution, but with colluding receivers this would not work since private keys could be shared in the collusion. 
\end{remark}

When classical communication is available, the receivers can collaborate to each send the $x_i$ amongst each other receiver to perfectly determine the channel state $s$ for each transmission. When all receivers are honest, nothing more needs to be done, however when some receivers are dishonest, more complexity is needed in order to preserve secrecy. We can turn to theory for secure multi-party computation (MPC), specifically secure multi-party modulo summation to compute the common function $\Phi(x^N)$. Since in this case the parties are not allowed to share entanglement and are allowed only classical communication, we only investigate purely classical strategies. In all strategies the assumption of minimum non-colluding parties (i.e. at most $N - 2$ parties can collude) is made. In the following we sketch the strategy and the complexity analysis.

\subsection{Strategy I: Secure multi-party modulo summation}

For the first strategy, we consider a protocol that requires secure channels between each of the receivers. We enforce this in this case in a purely classically way (e.g. using quantum key distribution schemes is prohibited). If we assume a standard public key cryptography system, then there is an additional communication overhead of $\Omega(N^2)$, since with just forward communication as the protocol needs,
\begin{align}
    (N-1)(N-2)/2 = (N^2-3N+2)/2,
\end{align}
secure channels are needed, and therefore as many public keys need to be distributed. Commonly used public key cryptography schemes like RSA are not unconditionally secure as they can be broken with enough computing power or a quantum computer \cite{shor1994algorithms}. In contrast, as we will see in Section \ref{sec:ent_cla}, when entanglement is available, secure communication is not necessary.

In \cite{chor1993communication}, Chor and Kushilevtiz give a multi-party protocol to compute a modulo sum with a security trade-off parameter $t \in [N]$. Here $t$ represents that no coalition of size at most $t$ can infer any additional information from the other receivers' $x_i$ from the other $N-t$ parties other than the modulo sum value, with $t \leq N-2$. This protocol uses $N\ciel{(t+1)/2}$ conferencing messages assuming preestablished secure communication channels. For completeness we state the protocol in Protocol \ref{proto:classical}, where $t$ is assumed to be $N-2$. 

\begin{protocol}[ht]
\caption{Classical State Decoding}
\begin{algorithmic}[1]
    \For{$i \in \{1,...,N-2 \}$}
        \State  \parbox[t]{313pt}{Receiver $i$ awaits $z_{j,i}$ from receivers $j < i$ and \\ calculates $w_i = \sum_{j=1}^{i-1} z_{j,i} \mod 2$ with $w_1 = 0$} 
        \State \parbox[t]{313pt}{Receiver $i$ selects uniformly at random \\ $z_{i,i+1}, z_{i,i+2}, ..., z_{i,N-1}$ each from $\{0, 1\}$ \vspace{1mm}}
        \State  \parbox[t]{313pt}{Receiver $i$ determines $z_{i,N}$ such that \\ $x_i + w_i = \sum_{j=i+1}^N z_{i,j}$  and sends $z_{i,j}$ to the respective \\ receiver $j>i$}
    \EndFor
    \State Receiver $N-1$ computes $z_{N-1,N} = x_{N-1} + \sum_{j=1}^{N-2}z_{j,N-1} \mod 2$ and sends it to receiver $N$
    \State Receiver $N$ computes $s =  x_N + \sum_{j=1}^{N-1} z_{j,n} \mod 2$ and broadcasts $s$ to all other receivers
\end{algorithmic}\label{proto:classical}
\end{protocol}

The proof of correctness for this protocol is given explicitly in \cite{chor1993communication}. The downside of using this protocol alone is that for decoding a message of length $n$, the protocol needs to execute $n$ times. In Strategy II, we use this protocol to generate a zero-sum random variable in the first step and then using this zero sum random variable, one can save resources by reusing it for each transmission. We explore the pros and cons in depth.

\subsection{Strategy II: Modulo-zero summation from modulo summation}

We consider another strategy where the receivers use Protocol \ref{proto:classical} to produce random variables $R = (r_1, ..., r_N)$ such that $\sum_{i=1}^N r_i = 0$ and then perform steps 4 and 5 from Protocol \ref{proto:summation} using this zero-sum randomness. To generate such a zero-sum random variable, one can use the protocol defined in \cite[Protocol 2]{hayashi2019verifiable} which we include explicitly as a subprotocol in Protocol \ref{proto:classic2}. Once a zero-sum realisation is generated, it can be used repeatedly to decode a message of length $n$. The full protocol in is given explicitly in Protocol \ref{proto:classic2}.

\begin{protocol}[ht]
\caption{Classical Secure Multi-Party Summation}
\begin{algorithmic}[1]
    \State Each receiver $i$ generates a uniform random number $Y_i$ independent of the other receivers
    \State The receivers run Protocol \ref{proto:classical} with $\{Y_i\}_{i=1}^N$ and each receiver will have the result $\sum_{i=1}^N Y_i$
    \State Receiver $1$ uses random variable $r_1 = Y_1 - \sum_{i=1}^N Y_i$, receivers $i\in\{2, ..., N\}$ use $r_i = Y_i$
    \State To determine channel state $s_t$ for transmission $1 \leq t \leq n$, each receiver broadcasts $s_{i,t} \coloneqq x_{i,t} + r_i$. The channel state is found by $s_t = \sum_{i=1}^N s_{i,t} \mod 2 = \sum_{i=1}^N x_{i,t} + r_i  \mod 2= \sum_{i=1}^N x_{i,t} \mod 2$ 
\end{algorithmic}\label{proto:classic2}
\end{protocol}

We summarize the communication complexity analysis. For establishing (conditionally) secure connections using public key distribution techniques requires each receiver (in the limit of increasing receivers) to send a public key to each other receiver consuming $\Omega(N^2)$ messages. Next, to generate a zero-sum random variable using Protocol \ref{proto:classical} requires $\Omega(Nt)$ for $t$ colluding parties. Once a zero-sum random variable realisation is successfully generated, the receivers can use step $4$ from Protocol \ref{proto:summation} which requires each sender use a broadcast channel to transmit their encrypted state information. In total, the communication is therefore $\Omega(N^2 + Nt)$.

That the zero-sum random variable can be recycled is because of its zero-sum condition. For example, assume $N-2$ parties collude and share their $r_i$ with each other in an attempt to determine, without loss of generality, $r_1$ and $r_2$. Then at the end of the protocol, the $N-2$ colluding parties have $(x_1 + r_1) + (x_2 + r_2) + \sum_{i=3}^N r_i = x_1 + x_2$, and thus no reminisce of $r_1$ or $r_2$ remains.

\section{Classical Communication with Entanglement}\label{sec:ent_cla}

\begin{theorem}\label{thm:comms}
When classical communication and entanglement is allowed between the receivers, there exists a multi-party summation protocol such the rate tuple $1^N$ is achievable with classical communication complexity $\Omega(N)$ and entanglement resource complexity $\Omega(N^2m)$ with unconditional security.
\end{theorem}

We sketch the proof. To determine the channel state $s$, the receivers perform Protocol \ref{proto:summation} for  multi-party modulo summation extended from  \cite[Protocol 4]{hayashi2019verifiable}, which offers verifiable randomness for zero-sum random variables and unconditional security. The protocol uses phase-GHZ states defined as follows. Let $R^N \coloneqq \{r^N = (r_1, ..., r_N) :\forall i, r_i \in \{0,1\}, \sum_{i=1}^N r_i = 0\}$. The phase-GHZ state is defined as:
\begin{align}
    \ket{GHZ}_p \coloneqq \frac{1}{\sqrt{2^{N-1}}}\sum_{r^N \in R^N} \ket{r^N}.
\end{align}
This ensures every joint measurement result $\ket{GHZ}_p$ sums to $0$. With security parameter $m$, we write the necessary protocols that the receivers use to securely communicate their piece of the channel state such that the security is verifiable.
\begin{protocol}[ht]
\caption{Phase-GHZ State Validation}
\begin{algorithmic}[1]
   \State With $4Nm$ copies of the $\ket{GHZ}_p$ state, randomly divide the states into $4m$ groups with $n$ states each. 
   \State Apply measurements to the respective groups using the measurement structure in \cite[Table 1]{hayashi2019verifiable}.
   \State Await a broadcast from all other receivers with their measurement results from the previous step.
   \State Check that inequalities $(9)$ and $(10)$ in \cite[Protocol 3]{hayashi2019verifiable} hold.
   \State If the inequalities hold report success, otherwise report failure.
\end{algorithmic}\label{proto: ghz validation}
\end{protocol}
\begin{protocol}[ht]
\caption{Entanglement Assisted State Decoding}
\begin{algorithmic}[1]
    \State The $N$ receivers generate $4N^2m + 1$ copies of the $\ket{GHZ}_p$ state.
    \State  From the $4N^2m + 1$ $\ket{GHZ}_p$ states, each receiver allocates $4N^2m$ of the copies, randomly selecting $4Nm$ copies from their respective part, and applies as a sub-protocol Protocol \ref{proto: ghz validation}. If the sub-protocol runs successfully, they continue to the next step.
    \State Each receiver $i$ measures the last remaining copy in the computational basis  and stores the output as $r_i$.
    \State For each $x_{i,t}$, each receiver $i$ calculates $s_{i,t} \coloneqq  x_{i,t} + r_i$ and broadcasts all $\{s_{i,t}\}_{t=1}^n$ to all other receivers.
    \State Each receiver calculates $s_t = \sum_{i=1}^{N} s_{i,t} \mod 2 = \sum_{i=1}^{N} x_{i,t} + r_i\mod 2 = \sum_{i=1}^{N} x_i \mod 2$. 
\end{algorithmic}\label{proto:summation}
\end{protocol}
Using these protocols, the receivers can with certainty and security (with respect to with parameter $m$) determine the channel state $s$ for each transmission, thereby allowing for perfect decoding the message.

The communication complexity of this approach is as follows. As far as we know, there are no protocols currently used for distributing GHZ states, and so to estimate the communication complexity of the task we assume there is a communal source generating $\ket{GHZ}_p$ states and transmitting them to the receivers in a \enquote{frame} of qubits using one transmission to transmit the $4N^2m + 1$ $\ket{GHZ}_p$ states, namely $\Omega(1)$ classical resources to generate $\Omega(N^2m)$ entanglement resources. Next, the receivers use Protocol \ref{proto: ghz validation} to validate their GHZ states using $\Omega(N)$ messages over a broadcast channel. Finally, the receivers again use  $\Omega(N)$ messages to broadcast the state information encoded with the zero-sum random variable. In summary, $\Omega(N^2m)$ entanglement resources are used with $\Omega(N)$ classical resources for verifiable unconditionally secure decoding of the state information satisfying all of the protocol requirements.

\section{Converse Arguments}

In each case, the converse arguments are the same. The broadcast channel transmits the messages in such a way that without signalling communication amongst the receivers, the channel acts as a $BSC(0.5)$ channel, well known to have a capacity of $0$. Because the channel state information vector $x^N$ is generated i.i.d. randomly, there is no correlation between any two receiver's individual state information. Because of this there is no way for the receivers to achieve any positive rate without the cooperation of all receivers. 

\section{Conclusions and Outlook}

In summary, we have constructed a communication scenario over a causal broadcast channel such that in order for the channel to have positive capacity, the receivers must cooperate to compute a multi-party modulo sum. We considered three scenarios where the receivers have access to various resources and showed that when the receivers can share entanglement and use a separate classical broadcast channel to cooperate, they can most efficiently determine the channel state to achieve the full capacity of the channel. We can further investigate different methods of selecting the channel state based results where there is a known quantum advantage.

\bibliographystyle{ieeetr}

\end{document}